\documentclass[reprint,aps,pra,amsmath,amssymb,twocolumn,final,floatfix,superscriptaddress,longbibliography,10pt]{revtex4-2}

\usepackage{amsmath}
\usepackage{algorithm}
\usepackage{graphicx} 
\usepackage{xcolor}
\usepackage[normalem]{ulem}
\usepackage[utf8]{inputenc}
\usepackage{mathtools}
\usepackage{braket}
\usepackage{mathtools}
\usepackage{hyperref} 
\hypersetup{
    colorlinks=true,       
    linkcolor=black,       
    citecolor=black,       
    filecolor=black,       
    urlcolor=black,        
    pdfborder={0 0 0},     
}

\begin{document}

\title{Limitations of Quantum Approximate Optimization in Solving Generic Higher-Order Constraint-Satisfaction Problems}

\author{Thorge Müller} \email{Thorge.Mueller@dlr.de} 
\affiliation{German Aerospace Center (DLR), Institute for Software Technology, Department High-Performance Computing, 51147 Cologne, Germany} 
\affiliation{Theoretical Physics, Saarland University, 66123 Saarbr{\"u}cken, Germany}

\author{Ajainderpal Singh} 
\affiliation{Institute for Quantum Computing Analytics (PGI-12),Forschungszentrum J{\"u}lich, 52425 J{\"u}lich, Germany} 

\author{Frank K. Wilhelm} 
\affiliation{Theoretical Physics, Saarland University, 66123 Saarbr{\"u}cken, Germany}
\affiliation{Institute for Quantum Computing Analytics (PGI-12),Forschungszentrum J{\"u}lich, 52425 J{\"u}lich, Germany}

\author{Tim Bode} 
\affiliation{Institute for Quantum Computing Analytics (PGI-12),Forschungszentrum J{\"u}lich, 52425 J{\"u}lich, Germany} 
\date{\today}
	
\begin{abstract}
The ability of the Quantum Approximate Optimization Algorithm (QAOA) to deliver a quantum advantage on combinatorial optimization problems is still unclear. Recently, a scaling advantage over a classical solver was postulated to exist for random 8-SAT at the satisfiability threshold. At the same time, the viability of quantum error mitigation for deep circuits on near-term devices has been put in doubt. Here, we analyze the QAOA’s performance on random Max-$k$XOR as a function of $k$ and the clause-to-variable ratio. As a classical benchmark, we use the Mean-Field Approximate Optimization Algorithm (MF-AOA) and find that it performs better than or equal to the QAOA on average. Still, for large $k$ and numbers of layers $p$, there may remain a window of opportunity for the QAOA. However, by extrapolating our numerical results, we find that reaching high levels of satisfaction would require extremely large $p$, which must be considered rather difficult both in the variational context and on near-term devices. 
\end{abstract}
	
\maketitle

\section{Introduction}

Quantum combinatorial optimization was applied to satisfiability problems early on~\cite{Hogg_2000}, underlining the naturalness of the question if, in essence, quantum superposition could aid combinatorial searches in exponentially large spaces. The interest in such questions only grew after the introduction of the QAOA by Farhi and Goldstone~\cite{farhi2014quantum}. Whether either adiabatic quantum computation~\cite{albashAdiabaticQuantumComputation2018} or the QAOA can deliver an advantage over any known best classical algorithm has remained an open question, however, not least because beating classical methods is a moving target.

Even so, Boulebnane and Montanaro~\cite{Boulebnane}, by analyzing the performance of the QAOA for the \textit{decision} problem $k$SAT with even $k$, recently found a potential speedup of the QAOA over a state-of-the-art classical solver when applied to random $8$SAT. A practical caveat to this result is the large number of QAOA layers required to push the (still exponential) scaling below the classical benchmark. A very recent result also hints at possible speedups for almost symmetric optimization problems from \textit{low-depth} QAOA~\cite{montanaro2024quantumspeedupssolvingnearsymmetric}.

This practical difficulty should not be underestimated, as a novel learning-theoretic proof~\cite{quekExponentiallyTighterBounds2024} has exponentially tightened the limitations of quantum error mitigation, thus severely reducing the prospects of variational quantum algorithms on near-term devices, which naturally includes the QAOA. One key message is that even shallow circuits will have a hard time estimating expectation values at current noise levels. In the worst case, this would mean that the QAOA is effectively limited to low numbers of layers until quantum error correction arrives. The QAOA had originally drawn attention~\cite{abbas2024} as one of the first quantum algorithms that might show an advantage over classical algorithms without fault-tolerant quantum computation. 

In the case of optimization problems involving higher-order Pauli-$Z$ strings, as explored below, the need to decompose these many-body interactions into two-body couplings further increases the circuit depth on platforms such as superconducting devices.

In a similar vein, a number of new results imply that models with a provable absence of barren plateaus turn out to be efficiently classically simulable~\cite{meleNoiseinducedShallowCircuits2024, cerezoDoesProvableAbsence2024}. Ref.~\cite{kaziAnalyzingQuantumApproximate2024} goes further by conjecturing that the Lie algebras of most problem graphs grow either exponentially or polynomially in the system size, meaning that the QAOA will either suffer from barren plateaus or turn out to be, yet again, classically simulable.

The interest in the QAOA's potential naturally extends beyond $k$SAT and their respective optimization problems, Max-$k$SAT, in particular to the related Max-$k$XOR problems, with Max-2XOR having received the most thorough scrutiny in the literature~\cite{Marwaha2021localclassicalmax,Fermionic_View,Fahri_II}. We remark that these problems belong to the complexity class NP-hard for $k \geq 2$~\cite{hasatd}. Compared to the non-exclusive OR $(\vee)$ of Max-$k$SAT, the literals per constraint of Max-$k$XOR problem instances are connected via an \textit{exclusive} OR $(\oplus)$. The conversion from a classical Max-$k$XOR instance to a quantum operator turns a $k$-literal constraint into a Pauli-$Z$ string of length $k$.

\begin{figure}[htb!]
	\begin{center}
    \includegraphics[width=\linewidth]{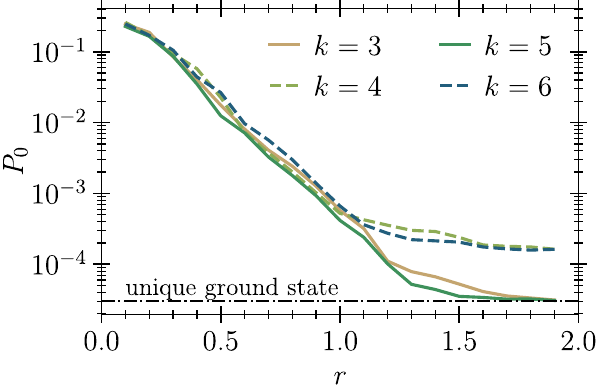}\vspace{-0.2cm}
    \caption{The normalized number of degenerate ground states, $P_{0}$ defined in Eq.~\eqref{eq:deg_gs}, against the clause-to-variable ratio $r$. We investigate $P_{0}$ for $k=3,4,5,6$ at $N=15$ and average over 400 instances per $k$ and $r$.}
    \label{fig:QAOA_KXOR_performance_r}
 	\end{center}
\end{figure}

Crooks~\cite{crooks2018performance} was one of the first to provide numerical evidence that the QAOA could indeed yield an advantage over classical algorithms on these problem types, showing that for an increasing number of parameters $2p$, the QAOA can at least surpass the $lower$ bound established by the classic Max-2XOR algorithm due to Goemans and Williamson~\cite{GW}, which derives from semi-definite programming (SDP). Note that Makarychev and Makarychev~\cite{Makarychev} showed that a general $k$-constrained satisfaction problem ($k$-CSP) has a lower bound than a $(k-1)$-CSP, i.e.\ increasing $k$ should be an interesting avenue to pursue in the search for an advantage.

Consequently, here we are delving into the performance of the QAOA on problems with $k>2$ literals (constraints) per clause, where Akshay et al.~\cite{qaoa_analytic_II} conducted one of the first performance investigations of the QAOA as a function of the clause-to-variable ratio $r$ {for Max-$k$SAT}. They showed that for $k=3$, one needs higher $p$ than for $k=2$ to achieve the same accuracy of the locally optimal solution. 

Another interesting result was given by Marwaha et al.~\cite{Marwaha} for \textit{uniform} Max-$k$XOR problems, who observed better performance than the best-known classical algorithm, the local threshold algorithm~\cite{threshold_algorithm}. The graphs associated with the problem instances in their investigation were \textit{regular} and \textit{triangle-free}. Similarly, the works of Chou et al.~\cite{Chou_ogp}, Chen et al.~\cite{chen_ogp}, and Basso et al.~\cite{Basso_2022} focused on random Max-$k$XOR problems where the related graph is \textit{sparse}. They especially drew attention to the so-called overlap gap property~\cite{gamarnik2019overlap,Gamarnik_2021} for even $k$; loosely speaking, this states that if two sub-optimal (or optimal) sets of satisfied clauses are close in energy, their bitstrings are either very close or very far from each other in Hamming distance. They showed analytically that, due to this property, the QAOA needs increasingly high $p$-values for Max-$k$XOR with even $k$. 

In contrast to these previous results, we explore \textit{arbitrary} random Max-$k$XOR problems. Additionally, we examine the performance specifically for \textit{odd} values of $k$, which complements the studies conducted in~\cite{Chou_ogp,chen_ogp,Basso_2022}.

This work is structured as follows: We first introduce the QAOA, the MF-AOA, and the Max-$k$XOR problem. Then we proceed by applying the QAOA and the MF-AOA to ensembles of randomly generated Max-$k$XOR instances. We begin by examining the dependence of their performance on the clause-to-variable ratio $r$. Subsequently, we analyze the impact of the parameter $k$ on the algorithms' behavior. Next, we explore the performance of the QAOA at high circuit depths, including an analysis of the distribution of the optimized parameters. Based on these results, we estimate the circuit depth at which QAOA achieves 99 \% approximation ratio. Finally, we compare the performance of the QAOA and MF-AOA across different values of $k$. All problem instances and the corresponding results are publicly available~\cite{data_paper}.

\section{Max-$k$XOR}

As mentioned, the distinction between $k$SAT and $k$XOR problems is that the variables within a clause of the latter are linked by the exclusive OR ($\oplus$) operator. In fact, a random $k$XOR decision problem can be written as
\begin{align}
C = \bigwedge_{\{i_1, i_2,.., i_k\} \in M} x_{i_{1}} \oplus x_{i_{2}} \oplus...\oplus x_{i_{k}},
\end{align}
where the $x_{1}, \ldots, x_{N} \in\{0,1\}$ are the $N$ variables defining the problem, and $M$ denotes the set of clauses. As an example, if we pick a 2XOR problem with three variables and three clauses, we have
\begin{align}
C = (x_1 \oplus x_2 ) \land (x_2 \oplus x_3  ) \land (x_1 \oplus x_3 ).
\end{align}
To go from the decision problem to the optimization problem, we transform $\wedge \rightarrow \sum$ and $\oplus \rightarrow+$. Hence, $C$ can also be expressed as
\begin{align}
\label{eq:C1}
C=\hspace{-3mm}\sum_{\left\{i_{1}, i_{2}, \ldots, i_{k}\right\} \in M} \hspace{-6mm} \left(x_{i_{1}}+x_{i_{2}}+\cdots+x_{i_{k}}+a_{i_{1} \ldots i_{k}}\right) \bmod 2,
\end{align}
where the symbol $a_{i_{1} \ldots i_{k}} \in\{0,1\}$ represents the parity sign corresponding to the selected clause, fixing the preferred truth convention for that specific clause. In the context of Max-2XOR, all parity signs are set to 1 and all variables are non-negated, resulting in the well-known standard formulation of Max-Cut. If instead we negate one variable in every clause by convention, each clause would be satisfied if and only if $both$ variables are equal, which is the opposite of standard Max-Cut.

The problem can be translated into a quantum operator $C \rightarrow \hat{C}$. To understand the transformation, we investigate the exemplary truth table of Max-3XOR in Tab. \ref{tab:truth_table_maxlin}. The outcome of the truth table (rightmost column of Tab.~\ref{tab:truth_table_maxlin}) can be represented by Pauli-$Z$ strings and the identity matrix. Hence, we identify $\{0,1\} \rightarrow\{1,-1\}$ and obtain $\hat{C}=1-\hat Z_{1} \hat Z_{2} \hat Z_{3}$ for this single-clause example. We can easily generalize this result to Max-$k$XOR, resulting in
\begin{align}
\label{eq:C2}
\hat{C}=\sum_{\left\{i_{1}, i_{2}, \ldots, i_{k}\right\} \in M} \frac{1}{2}\left(1 \pm \hat Z_{i_{1}} \hat Z_{i_{2}} \cdots \hat Z_{i_{k}}\right),
\end{align}
where the signs depend on the respective parities $a_{i_{1} \ldots i_{k}}$ from Eq.~\eqref{eq:C1}.

\begin{figure}[htb!]
    \centering
    \includegraphics[width=\linewidth]{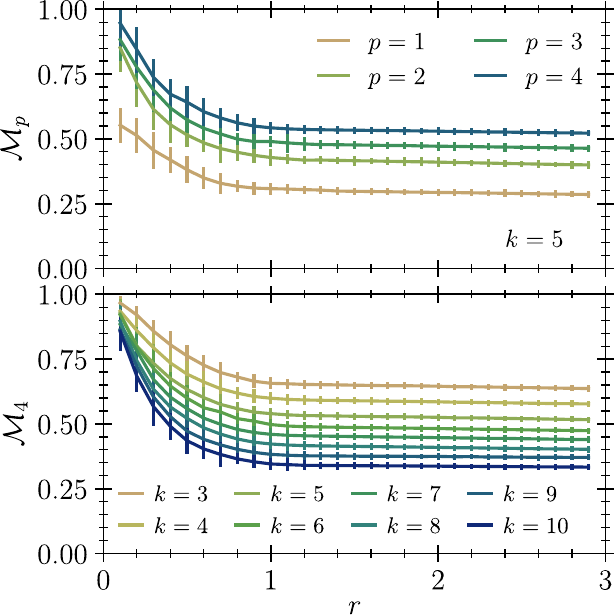}\vspace{-0.2cm}
    \caption{Ensemble-averaged approximation ratio of the QAOA at $N=18$. For every $r, k, p$, we average over 10000 random instances. (Upper panel) $\mathcal{M}_{p}$ against the clause-to-variable ratio $r$ for $k=5$ and $p=1,2,3,4$. (Lower panel) $\mathcal{M}_{p}$ as a function of the parameter $r$ for $k \in[3,10]$.}
    \label{fig:M_p_vs_r}
\end{figure}

We finish this section by discussing how to sample these random Max-\textit{k}XOR instances. The clause-to-variable ratio is defined as
\begin{align}
\label{eq:Ratio}
r=\frac{|M|}{N}.
\end{align}
Because we are interested in \textit{random} graphs, we need to determine the clause probability corresponding to specific values of $r, N$ and $k$. Now, one might assume that there are $2^{k}$ total clauses for a given $k$-literal, given the inclusion of negated variables. This would imply that the number of all possible clauses for Max- $k$XOR at system size $N$ is $2^{k}\binom{N}{k}$. However, by looking at the exemplary truth table shown in Tab. \ref{tab:truth_table_maxlin}, we see that the truth value of each of the $2^{k}$ clauses only depends on the \textit{parity} of negations, i.e.\ an \textit{odd} number of negated variables results in the corresponding clause being violated, while an even number of negations results in a true clause. Effectively, then, for the $k$-literal of Tab. \ref{tab:truth_table_maxlin}, we have only \textit{two} distinct outcomes with \textit{opposite} parity signs in the cost function. Clearly, including both of these would render the cost function trivial. For this reason, we randomly choose among these two possibilities for each $k$-literal, which results in {2}$\binom{N}{k}$ being the effective total number of clauses. Therefore, to find a specific random instance of Max-$k$XOR for given $r$, one can pick randomly from the set of all possible clauses with a probability
\begin{align}
\mathcal{P}= r N \binom{N}{k}^{-1}.
\end{align}
To calculate the performance of QAOA applied to a specific Max-$k$XOR problem with fixed $r$, we then average over a set of random instances per value of $k$. Furthermore, we only take pure $k$-literal constraints into account (instead of $(k-1)$-literals etc.) because we want to investigate the performance of the QAOA in clean dependence on $k$.

\begin{table}[htb!]
	\centering
	\begin{tabular}[t]{c|c|c|c}
		$x_1$ & $x_2$ & $x_3$ & $x_1$ $\oplus$ $x_2$ $\oplus$ $x_3$ \\
		\hline
		0 & 0 & 0 & 0\\
		0 & 0 & 1 & 1\\
		0 & 1 & 0 & 1\\
		0 & 1 & 1 & 0\\
		1 & 0 & 0 & 1\\
		1 & 0 & 1 & 0\\
		1 & 1 & 0 & 0\\
		1 & 1 & 1 & 1\\
	\end{tabular}
	\caption[7]{ Truth table for a 3XOR clause consisting of the $N=3$ variables $x_{1}, x_{2}$ and $x_{3}$. The rightmost column shows the results under which assignment to the variables the clause is satisfied. }
	\label{tab:truth_table_maxlin}
\end{table}%

\section{Methods}

\subsection{The QAOA}

Every constrained satisfaction problem with $k$ literals can be converted into a quantum operator consisting of diagonal quantum Pauli-$Z$ strings, expressed as
\begin{align}
\label{eq:Problem}
\hat{C} & =\sum_{i_{1}, \ldots, i_{k}} J_{i_{1}, \ldots, i_{k}} \hat Z_{i_{1}} \cdots \hat Z_{i_{k}}, 
\end{align}
where the $J_{i, \ldots, i_{k}}$ are real. The corresponding unitary operator is given by
\begin{align}
\hat U_C(\gamma)=\exp (-\mathrm{i} \gamma \hat{C}).
\end{align}
\begin{figure}[htb!]
	\centering
  \includegraphics[width=\linewidth]{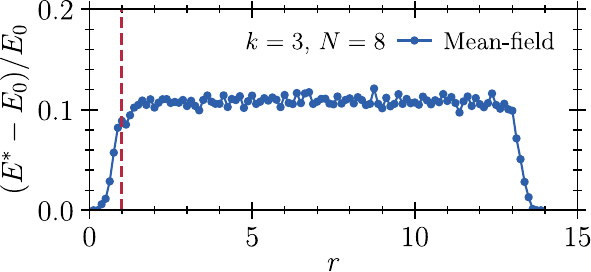}\vspace{-0.2cm}
	\caption{The relative energy deviation $(E^*-E_0)/E_0$ plotted as a function of $r$ for $k=3$ and $N=8$, covering the entire range of possible $r$. Each point represents the average over 500 randomly generated instances. The dotted line indicates the phase transition from easy to hard for the $k$XOR decision problem.}
	\label{fig:complete}
\end{figure} 
In the QAOA, the cost function or problem Hamiltonian is complemented by a so-called \textit{driver} Hamiltonian typically inducing a unitary
\begin{align}
\label{eq:driver}
\hat U_{X}(\beta)=\exp \left(-i \beta \sum_{i=1}^N \hat X_{i}\right) 
\end{align}
comprised of single-qubit $X$ rotation gates. These gates, often called transverse-field operators in the quantum-annealing literature~\cite{transverse_field}, are responsible for driving transitions between computational basis states, i.e. solution states. The QAOA ansatz is then constructed from $p$ pairs of these unitaries,
\begin{align}
|\gamma, \beta\rangle =\hat U_{X}\left(\beta_{p}\right) \hat U_C\left(\gamma_{p}\right) \cdots \hat U_{X}\left(\beta_{1}\right) \hat U_C\left(\gamma_{1}\right) |+\rangle^{\otimes n}, 
\end{align}
where $\gamma, \beta$ are classical parameters. From the ansatz state, the expectation value of the cost function is computed as
\begin{align}
F_{p}(\gamma, \beta)=\langle\gamma, \beta| \hat{C}|\gamma, \beta\rangle,
\end{align}
which serves as the loss function to be optimized, i.e.\ the aim is to determine the values of $\gamma$ and $\beta$ that maximize $F_{p}(\gamma, \beta)$. To quantify the performance of the QAOA, we introduce the approximation ratio
\begin{align}\label{eq:M_p}
\mathcal{M}_{p}=\frac{\max_{\gamma, \beta} F_{p}(\gamma, \beta) - E_{\min }}{E_{\max }-E_{\min}},
\end{align}
where $E_{\min }$ and $E_{\max }$ are the minimum and maximum values of $\hat{C}$, respectively.

\subsection{Mean-Field Approximate Optimization}

The MF-AOA~\cite{MeanField} can be viewed the classical counterpart to the QAOA, where quantum evolution is replaced by classical spin dynamics via the mean-field approximation. A simplified version of it has indeed been considered in the early days of quantum optimization~\cite{smolinClassicalSignatureQuantum2014, shinHowQuantumDWave2014}. Here, we generalize the algorithm to arbitrary $k$-spin interactions as occurring in $k$-CSPs. Through  the analogy of the fixed-angle QAOA with adiabatic quantum computation, we introduce the adiabatic Hamiltonian as 
\begin{align}\label{eq:H_adab}
    \hat H(s) = (1 - s(t)) \sum_{i=1}^N \hat X_i + s(t) \hat C,
\end{align}
where $s(t) = t/T_f$, $s(t) \in [0, 1]$ is the scaled time. The mean-field form of this Hamiltonian is then given by
\begin{align}
\label{eq:mf_problem}
    \begin{split}
        H(s) &= \left(1 - s(t)\right) \sum_{i=1}^N  n^x_i(t) \\
        &+ s(t)\sum_{i_{1}, \ldots, i_{k}} J_{i_{1}, \ldots, i_{k}} n^z_{i_{1}}(t) \cdots n^z_{i_{k}}(t) , 
    \end{split}
\end{align}
with $\boldsymbol{n}_i(t) = (n^x_i(t), n^y_i(t), n^z_i(t))^T$ being the classical spin vectors living on their respective Bloch spheres. The effective magnetization
\begin{align}
    \begin{split}
        m_{i_j}(t) &=\sum_{i_{1}, \ldots, i_{j-1}, i_{j+1}, \dots, i_{k-1}}\hspace{-2mm} J_{i_{1}, \ldots, i_{j-1}, i_{j+1}, \dots, i_{k-1}} \\
        &\times n^z_{i_{1}}(t)\cdots n^z_{i_{j-1}}(t)n^z_{i_{j+1}}(t)\cdots n^z_{i_{k}}(t)  
    \end{split}
\end{align}
then determines the classical equations of motion as 
\begin{align}\begin{split}
\dot n^x_i(t) &= -2 s(t) m_i(t) n^y_i(t),\\
\dot n^y_i(t) &= \phantom{-}2 s(t) m_i(t) n^x_i(t) - 2 (1 - s(t))  n^z_i(t) , \\
\dot n^z_i(t) &= \phantom{-}2 (1 - s(t))  n^y_i(t).
\end{split}\end{align}
This set of $3N$ non-linear ordinary differential equations can be solved numerically in an efficient way~\cite{Tim}. These equations can be derived, e.g., from the Heisenberg equations of motion via a product ansatz for the density matrix. We also remark that a similar product ansatz is used to derive near-optimal approximation ratios for the quantum generalization of Max-Cut in Ref.~\cite{gharibian_2019}.

For $\mathbb{Z}_2$-symmetric problem Hamiltonians, i.e. {those} without a \textit{local} tensor $J_{i_1}$, the effective magnetization evidently remains zero throughout. To circumvent this issue, in the MF-AOA it is necessary to explicitly break the $\mathbb{Z}_2$ symmetry. A useful way of achieving this is to add a so-called catalyst Hamiltonian~\cite{ghosh2024exponentialspeedupquantumannealing}
\begin{align}\label{eq:H_cat}
    \hat H_{\mathrm{cat}} =  \sum_{i=1}^N \Lambda_i(t) \hat Z_i
\end{align}
to Eq.~\eqref{eq:H_adab}, where the coefficients $\lambda_i(t)$ need to obey the boundary conditions $\lambda_i(0) = \lambda_i(T_f) = 0$. For our purposes, it suffices to take $\Lambda_i(t) = \lambda_i s^2 (t)(1 - s(t))$, where the $\lambda_i$ are chosen from a normal distribution $\mathcal{N}(0, \sigma^2)$. For the purpose of this paper, the standard deviations $\sigma$ are chosen according to Table~\ref{tab:sigmas}.

\begin{table}\label{tab:sigmas}
\centering
\begin{tabular}{c|cccccccc}
$k$ & 3 & 4 & 5 & 6 & 7 & 8 & 9 & 10 \\
\hline
$r=0.5$ & 0.2 & 0.5 & 1.0 & 1.0 & 1.5 & 1.5 & 2.0 & 3.0 \\
\hline
$r=1.5$ & 0.5 & 1.0 & 1.0 & 1.0 & 1.5 & 1.5 & 1.5 & 3.0 \\
\end{tabular}
\caption{Standard deviations $\sigma$ used in the random catalysts of Eq.~\eqref{eq:H_cat} for different $k$ and $r$.}
\end{table}

\begin{figure*}[htb!]
	\centering
  \includegraphics[width=\linewidth]{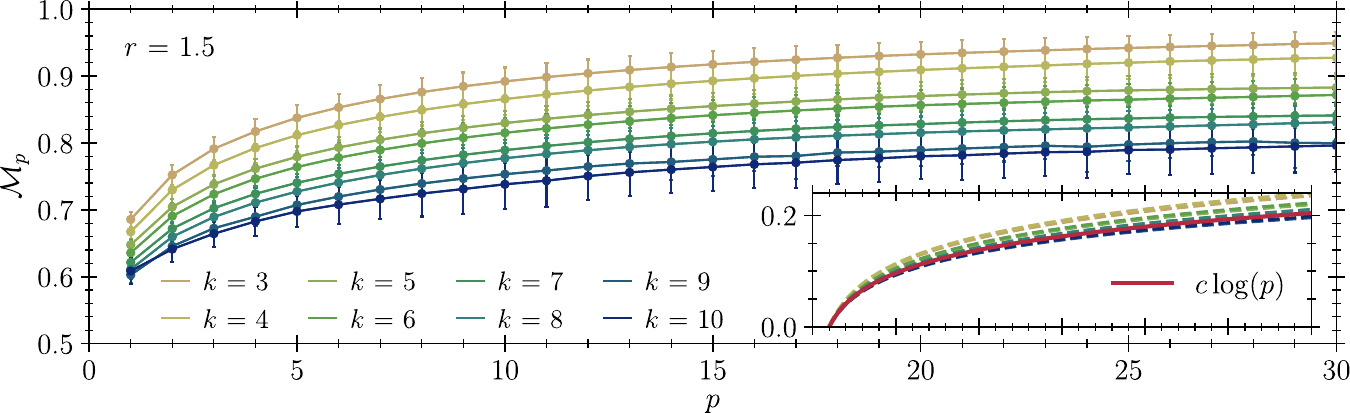}\vspace{-0.2cm}
	\caption{The ensemble-averaged approximation ratio $\mathcal{M}_{p}$ against $p$ for Max-$k$XOR with $k \in\{3, \ldots, 10\}$. We set $N=18$ and $r=1.5$ to investigate hard-to-solve instances for the QAOA. For each $p$ and $k$, we average over 100 random instances. The inset shows the collapsed logarithmic fits to the data (i.e.\ subtracting each constant shift). The solid (red) line is drawn for illustration with a coefficient of $c=0.06$.}
	\label{fig:kxor_p}
\end{figure*} 

Note that this additional Hamiltonian term does not affect the initial and final spectrum, as it is zero at both the beginning and the end of the adiabatic evolution. The effective magnetization can now be written as 
\begin{align}
    \begin{split}
        m_{i_j}(t) &= \Lambda_{i_j}(t) + \sum_{i_{1}, \ldots, i_{j-1}, i_{j+1}, \dots, i_{k-1}}\hspace{-2mm} J_{i_{1}, \ldots, i_{j-1}, i_{j+1}, \dots, i_{k-1}} \\
        &\times n^z_{i_{1}}(t)\cdots n^z_{i_{j-1}}(t)n^z_{i_{j+1}}(t)\cdots n^z_{i_{k}}(t).  
    \end{split}
\end{align}
The equations of motion are then evolved adaptively up to a final time $T_f=2^{15}$ in units of the inverse initial transverse field, and the solution bit string is obtained by projecting the $z$-components of the final spin vectors $\boldsymbol{n}_i(T_f)$ to the poles of their Bloch spheres via
\begin{align}
    \boldsymbol{\sigma} = (\mathrm{sign}(n_1^z(T_f)), \ldots, \mathrm{sign}(n_N^z(T_f))^T.
\end{align}

\begin{figure}[htb!]
	\centering
  \includegraphics[width=\linewidth]{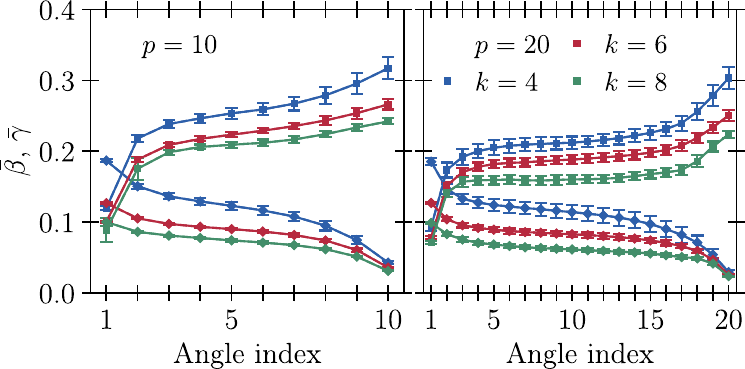}\vspace{-0.2cm}
	\caption{The ensemble-averade optimal parameters $\bar{\beta}, \bar{\gamma}$ for the QAOA with $p=10$ (left plot) and $p=20$ (right plot). The averages are taken over the ensembles of Figure~\ref{fig:kxor_p} with 100 instances per $k$ and $p$.}
	\label{fig:parameters}
\end{figure}

\section{Results}
\subsection{QAOA approximation ratio as a function of the clause-to-variable ratio}

The clause-to-variable ratio $r$ is a critical property of specific instances of CSPs. As the clause-to-variable ratio increases, they transition from an under-constrained regime to an over-constrained regime.
Generally speaking, over-constrained instances will be harder than under-constrained ones. This statement should be understood in a statistical sense: at high $r$, hard random instances will be more prevalent, while easy instances may still occur.

For small $r$, the instances are under-constrained, and typically not every variable appears in every clause; hence multiple solutions can exist. Such instances are easy to solve for a given algorithm. Additionally, the effective problem size $N$ can be reduced. To construct under-constrained and over-constrained problems, we define
\begin{align}\label{eq:deg_gs}
P_{0}=2^{-N} \sum_{i} E_{0, i} 
\end{align}
as the ratio of the number of degenerate ground states and the total number of possible variable assignments, which is $2^{N}$. Figure~\ref{fig:QAOA_KXOR_performance_r} illustrates how $P_{0}$ scales with $r$, e.g. for small $r=0.1$, we have $P_{0} \approx 0.15$, which means that $15 \%$ of all states correspond to the optimal solution, making these instances highly under-constrained. As $r$ increases, the number of optimal solutions drops quickly. For $r=1.5$, we already have $P_{0} \approx 10^{-5}$ for odd $k=3,5$ and $P_{0} \approx 10^{-4}$ for even $k=4,6$. For $k=3,5$ in particular, the instances approach unique solutions for high $r$. 
\begin{figure}[htb!]
	\centering
  \includegraphics[width=\linewidth]{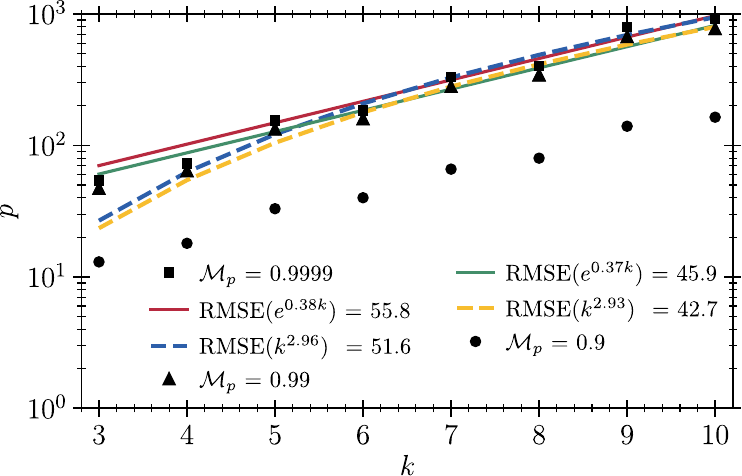}\vspace{-0.2cm}
	\caption{The required $p$ to reach various ensemble-averaged approximation ratio thresholds $\mathcal{M}_p$ as a function of $k$. The value of $p$ is determined by fitting a logarithmic function to Figure~\ref{fig:kxor_p}. The dotted curves represent exponential and polynomial fits, along with the Root-Mean-Squared Error (RMSE) of each fit.}
	\label{fig:99_percent}
\end{figure} 
Such instances with only a few ground states are over-constrained because every variable is typically included in \textit{at least} one clause. An exact threshold for under- and over-constrained random instances does not exist, but as a rule of thumb, we can state that $r<1$ corresponds to less than one clause per variable. We also note that the decrease in the approximation ratio slows down for all $p$ and $k$ at high values of $r$.

The upper panel of Figure~\ref{fig:M_p_vs_r} displays the results for the QAOA applied to Max-$k$XOR as a function of $r$. We fix $k=5$ and investigate the ensemble-averaged approximation ratio for $p=1,2,3,4$; we observe a clear dependence on \textit{r}. Crucially, $\mathcal{M}_{p}$, as defined in Eq.~\eqref{eq:M_p}, exhibits a significant decrease with growing $r$, regardless of the specific Max-$k$XOR problem and the value of $p$. For high $r$, the instances become hard to solve. Importantly, higher values of $p$ merely result in a vertical \textit{shift} along the $\mathcal{M}_{p}$-axis and \textit{not} in a systematically improved scaling. We remark that increasing $r$ until all-to-all connected problem graphs are reached would lead to another under-constrained regime in which the approximation ratio of the QAOA should improve again. The Max-5XOR problem is used as a representative; we did not observe qualitatively different results for other values of $k$.

\subsection{Mean-field approximation ratio as a function of the clause-to-variable ratio}

To investigate how the mean-field approximation ratio depends on the clause-to-variable ratio $r$, we apply the MF-AOA to random instances of the Max-$k$XOR problem across the entire range of possible 
$r$ values for fixed $N$ and $k$. For this problem, the number of clauses ranges from 1 to  $2{N\choose k}$. We plot the relative energy deviation $(E^*-E_0)/E_0$ as a function of $r$, as this metric quantifies how close the system is to the true ground state of the problem Hamiltonian. It also serves as a useful hardness measure for MF-AOA across different problem instances.

\begin{figure}[htb!]
	\centering
  \includegraphics[width=0.95\linewidth]{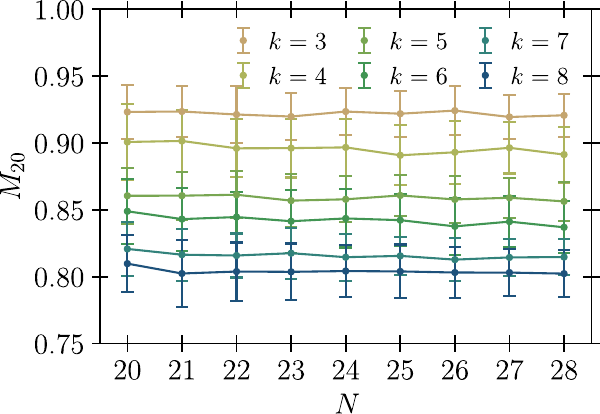}\vspace{-0.2cm}
	\caption{Ensemble-averaged approximation ratio $\mathcal{M}_{20}$ against system size $N$. Per $N$ and Max-$k$XOR we average over 100 instances.}
	\label{fig:QAOA_XOR_N}
\end{figure}

In the decision variant of the $k$-XOR problem, a well-known phase transition occurs at 
$M=N$, corresponding to $r=1$. Instances with $r<1$ or $r \gg 1$ are computationally less challenging to solve compared to those around $r=1$. For the optimization variant, the results presented in Figure~\ref{fig:complete} demonstrate distinct performance regimes. In the under-constrained regime ($r<1$), where the associated problem graph is sparse, the system remains closer to the ground state. Conversely, in the hard regime ($r>1$) the ensemble-averaged approximation ratio plateaus, becoming largely independent of the clause-to-variable ratio.

Since the approximation ratio in the hard regime does not vary significantly with $r$, we fix $r=0.5$ for the easy case and $r=1.5$ for the hard case in the remainder of the paper.

\subsection{QAOA approximation ratio as a function of $k$}

\begin{figure}[htb!]
	\centering
  \includegraphics[width=\linewidth]{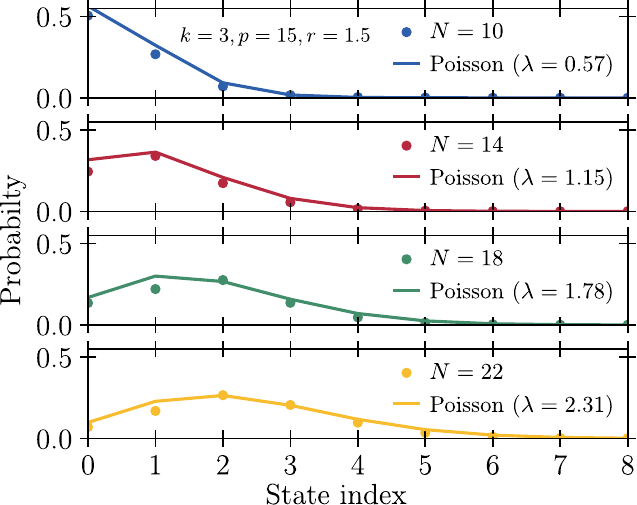}\vspace{-0.2cm}
	\caption{Concentration of the ensemble-averaged probability distributions of the final states reached by the QAOA for $k=3$. The state indices are ordered by energy,  with 0 indicating the ground state. Above the 8th excited state, the probabilities become negligible. Note that the Poisson distributions are \textit{not} fitted but simply evaluated at the mean state index.}
	\label{fig:QAOA_XOR_N_Poisson}
\end{figure}

In this section, we investigate how the ensemble-averaged approximation ratio of the QAOA changes when applied to random Max-$k$XOR instances for large $k$ and fixed $p=4$. The results are shown in the lower panel of Figure~\ref{fig:M_p_vs_r}. For large $r$, we observe that $\mathcal{M}_{4}$ \textit{decreases} with $k$, providing a first indication that random Max-$k$XOR becomes harder and harder to solve for the QAOA. This is in contrast to the results of Marwaha et al.~\cite{Marwaha} for uniform instances, who observed an \textit{increase} in the QAOA's approximation ratio with $k$. In Figure~\ref{fig:M_p_vs_r}, it is evident that the standard deviation is larger for small $r$ than for large $r$ for all values of $k$. This is due to a higher variability in the difficulty of random instances when $r$ is closer to zero. We also {observe}  the impact of \textit{delocalization} on the QAOA. To illustrate the delocalization effect, we choose a constant value for $k$. Upon selecting a random assignment of the variables for a given instance, flipping a single bit induces a change in the cost function. If we now calculate this change of cost for an instance with $k+1$, we observe an \textit{increase} as compared to $k$. The QAOA is of local structure through its single-qubit $X$-driver, leading to a decrease in the ensemble-averaged approximation ratio as $k$ increases.

\subsection{QAOA approximation ratio at high circuit depth}

This section analyzes how the QAOA performs on hard-to-solve random instances at large $p$. As previously discussed, solving {random} instances with large $r$ is more difficult than those with small $r$, regardless of the value of $k$, unless the problem-related graph is nearly all to all connected, i.e.
\begin{align}
r \approx\binom{N}{k} \cdot N^{-1}.
\end{align}
In Figure~\ref{fig:kxor_p}, we plot the results for $\mathcal{M}_{p}$ with $p \in\{1, \ldots, 30\}$ and $k \in\{3, \ldots, 10\}$. As $k$ increases, solving Max-$k$XOR problems with the QAOA becomes significantly more challenging. This trend suggests that an exponentially larger number of layers is required to reach the ground state for high values of $k$. Extending the findings of Basso et al. [15], who demonstrated the inefficiency of the QAOA for solving Max-$k$XOR problems with \textit{even} values of $k$, we thus provide numerical evidence that this assertion seems to hold for \textit{all} values of $k$.

\begin{figure}[htb!]
	\centering
  \includegraphics[width=\linewidth]{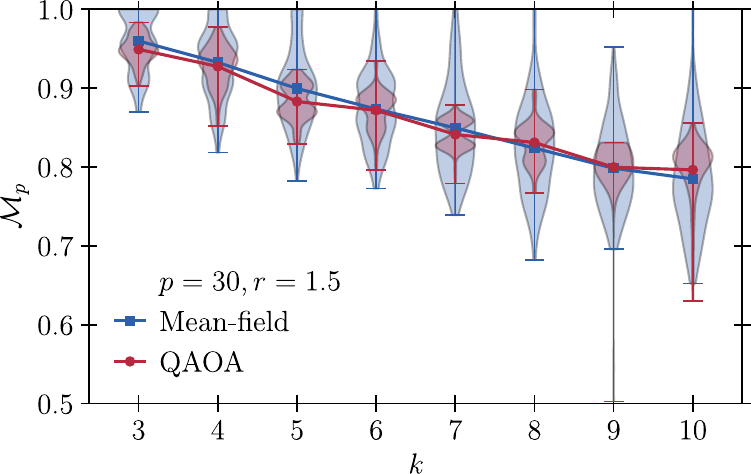}\vspace{-0.2cm}
	\caption{The distribution of $\mathcal{M}_{p}$ against $k$ for Max-$k$XOR with $k \in\{3, \ldots, 10\}$. We set $N=18$ and $r=1.5$ to compare hard-to-solve instances for the QAOA against mean-field. While the shaded area indicates the outcome distribution, the points represent averages over 100 random instances.}
	\label{fig:Comp_k}
\end{figure}

To assess the limitations of the QAOA, we estimate the number of layers $p$ required to an approximation ratio of {reach 90\% and above} for challenging problem instances in Figure~\ref{fig:99_percent}. Specifically, we fit the data from Figure~\ref{fig:kxor_p} to a logarithmic ansatz (see inset of Figure~\ref{fig:kxor_p}). To improve the accuracy of the fit for higher $p$ values, we exclude the first data point ($p=1$). By extrapolating the resulting fits to high $p$, we can estimate the $p$-value at which a given $\mathcal{M}_{p}$ is achieved. For lower values of $k$ (e.g., $k=3$), a circuit depth of $p \approx 50$ is required to reach 99\% of the ground-state energy. For larger $k$ (e.g., $k = 10$), this increases to approximately $p = 770$ layers. These findings highlight the limitations of the QAOA on current hardware, where gate fidelities of around 99.5\% make it impractical to implement circuit depths as large as $p = 50$.

To run QAOA for higher values of 
$p$, we employ a random initialization strategy with $1000$ random starting points and use the COBYLA optimizer for $p \leq 3$. For $p > 3$, we transition to the linear-interpolation strategy developed by Zhou et al.~\cite{qaoa_analytic_sim}. Figure~\ref{fig:parameters} presents the average results for the calculated parameters. The standard deviation remains low for small values of $p$ and increases for higher $p$. Beyond $p=30$, the linear-interpolation method shows no further improvement. In contrast, the Fourier-based approach proposed by Zhou et al.~\cite{qaoa_analytic_sim} demonstrates promising performance for $p>30$, albeit at a significantly higher computational cost.

\subsection{QAOA approximation ratio at larger $N$} 

In the previous chapters, we set $N=18$ to be able to simulate the QAOA with reasonable computational effort. In this section, we now aim to investigate how the QAOA performs for larger system sizes $N$. To compute the values of $\mathcal{M}_p$, we use the optimal angles from Figure~\ref{fig:parameters} found at $N=18$ and apply these parameters to instances with $N>18$ for each $k$, i.e.\ we do not perform the parameter optimization for $N>18$. Exemplary results for $p=20$ are shown in Figure~\ref{fig:QAOA_XOR_N}.

\begin{figure}[htb!]
	\centering
  \includegraphics[width=\linewidth]{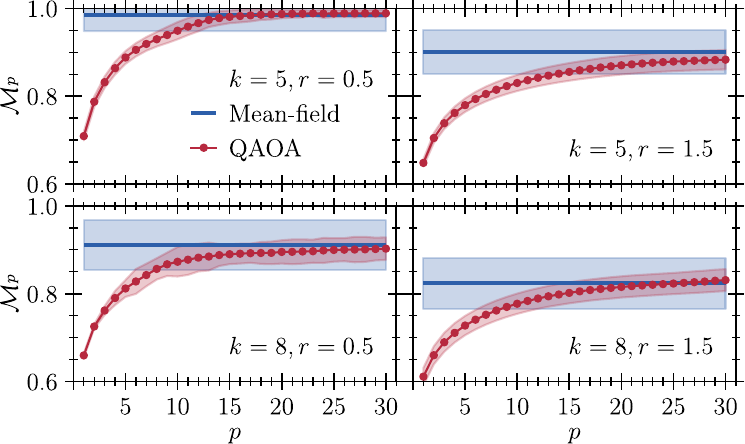}\vspace{-0.2cm}
	\caption{Ensemble-averaged approximation ratio $\mathcal{M}_{p}$ against $p$ for the QAOA (solid line) and the MF-AOA (dashed horizontal line) across four different combinations of $k$ and $r$. Each line represents an average over 100 random instances, with the shaded areas indicating the standard deviation.}
	\label{fig:Comp_p}
\end{figure} 

We observe that, for each $k$, the ensemble-averaged approximation ratio of the QAOA remains nearly constant across all investigated values of $N$, suggesting a certain degree of universality of our optimized angles. These findings are consistent with those of Farhi et al.\cite{Fahri_II}, who demonstrated that parameters originally derived for the infinite-size Sherrington-Kirkpatrick model were also applicable to finite-size instances.

To investigate the transition in behavior from small ($N \approx 10$) to larger system sizes ($N \approx 20$), we analyze the ensemble-averaged probability distribution of the QAOA across states ranging from the ground state to the 8th excited state. These distributions, plotted in Figure~\ref{fig:QAOA_XOR_N_Poisson} for $k=3$, reveal a trend that closely follows a Poisson distribution for different values of $N$. For small systems the distribution exhibits concentration of states near the ground state. As $N$ increases, this concentration diminishes, accompanied by a noticeable shift towards higher excited states.

\subsection{MF-AOA vs. QAOA}

In this section, we compare the ensemble-averaged approximation ratio of the QAOA with the MF-AOA by analyzing $\mathcal{M}_{p}$ across random Max-$k$XOR instances. This provides insight into how both algorithms compare across varying problem complexities and circuit depths.

\begin{figure}[htb!]
	\centering
  \includegraphics[width=\linewidth]{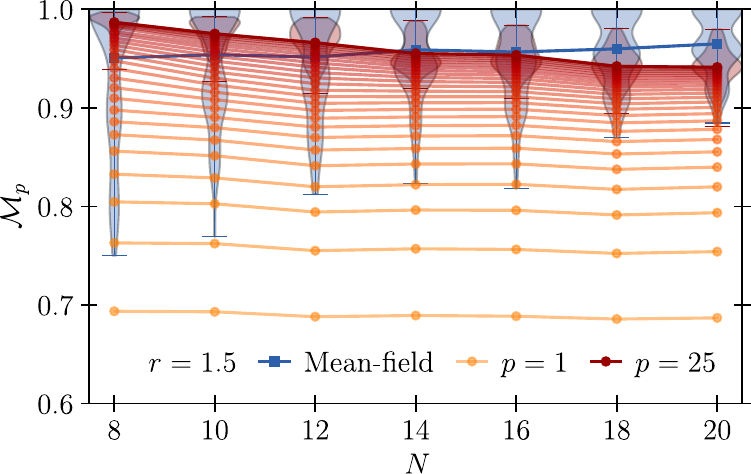}\vspace{-0.2cm}
	\caption{Ensemble-averaged approximation ratio $\mathcal{M}_{p}$ against system size $N$ for the MF-AOA and the QAOA ranging from $p=1$ to $p=25$ for $k=3$. Each line represents an average over 100 random instances. The clearly visible convergence of the QAOA's approximation ratio with increasing $p$ raises the question whether anything is to be gained by going to deeper circuits.}
	\label{fig:Comp_N}
\end{figure}

In Figure~\ref{fig:Comp_k}, we plot the distribution of $\mathcal{M}_{p}$ as a function of $k$, with a fixed $r=1.5$ and circuit depth $p=30$ for the QAOA. This allows us to compare both algorithms' ensemble-averaged approximation ratio for increasing values of $k$, which corresponds to the number of literals in each clause of the Max-$k$XOR problem. We observe that the MF-AOA exhibits a broadened distribution of $\mathcal{M}_{p}$, indicating greater variance of the MF-AOA across random instances. Despite the wider distribution, the MF-AOA consistently matches the QAOA on average. Notably, the QAOA’s ensemble-averaged approximation ratio relative to MF-AOA improves as $k$ increases. It seems likely that both the increased MF-AOA variance and the slightly lower approximation ratio at larger $k$ can be addressed by sampling more random catalysts according to Eq.~\eqref{eq:H_cat}.
{To identify a region where the QAOA outperforms its classical counterpart (MF-AOA), we compare both algorithms in Figure~\ref{fig:Comp_N} across varying system sizes and circuit depths. Our findings show that for small systems ($N \approx 8$), the QAOA outperforms the MF-AOA. However, as the system size increases ($N \approx 20$), the MF-AOA surpasses the QAOA.  Interestingly, for small values of $p$, the QAOA's approximation ratio remains nearly independent of the system size.}

In Figure~\ref{fig:Comp_p}, we plot $\mathcal{M}_{p}$ against $p$ for four distinct combinations of $k$ and $r$, representing both easy and hard cases, as well as low and high values of $k$. We compare the ensemble-averaged approximation ratio of the QAOA (in dependence on the circuit depth) with that of the MF-AOA. For hard instances ($r=1.5$), we observe that the QAOA approaches and eventually surpasses the MF-AOA's approximation ratio as the circuit depth increases. However, the QAOA requires relatively high circuit depths to match the $\mathcal{M}_{p}$ value achieved by the classical benchmark for both low and high $k$ values. Notably, for the easy case($r=0.5$) at $k=5$, the QAOA matches the classical approximation ratio at relatively low circuit depth $p=17$ and even surpasses it at large depths. However, the MF-AOA results require hardly any computational effort, while the angle optimization going into the QAOA results consumes considerable resources (besides being problematic on noisy quantum hardware~\cite{quekExponentiallyTighterBounds2024, meleNoiseinducedShallowCircuits2024, cerezoDoesProvableAbsence2024, kaziAnalyzingQuantumApproximate2024}). 

\section{Conclusion \& Outlook}

Our numerical analysis demonstrates that the \textit{relative} improvement of the QAOA's ensemble-averaged approximation ratio applied to Max-$k$XOR declines as both the parameter $k$ and the number of layers $p$ increase. Our results provide explicit numerical evidence detailing the approximation ratio of the QAOA across different Max-$k$XOR instances (for odd and even $k$) and tentatively predict the circuit depth required to achieve an approximation ratio $\mathcal{M}_p$ exceeding $90 \%$. Additionally, we observe that the parameters optimized for a fixed problem size ($N=18$) appear to generalize well to larger systems. These results suggest that achieving solutions close to optimal will require significantly higher circuit depths than presently available. These findings align with the analytic observations of Basso et al.~\cite{Basso_2022}, who noted challenges in using the QAOA to approximate ground state energies for sparse Max-$k$XOR problems (even $k$), especially for larger $k$, with shallow circuits.

Interestingly, the same approximation-ratio decrease in $k$ is experienced by our classical benchmark, seemingly resulting in a slight average advantage for the QAOA at larger $k$, see Figure~\ref{fig:Comp_k}. It would be intriguing for future work to look into the behavior of other classical solvers in the same limit.
We also point out that our mean-field results should be considered a \textit{lower bound} on the possible ensemble-averaged approximation ratio, as there is, at the very least, the option to sample more instantiations of our random catalyst from Eq.~\eqref{eq:H_cat}. Since any proper time-to-solution analysis~\cite{albashAdiabaticQuantumComputation2018} has to take into account optimizing the QAOA angles, there is a lot of leeway for classical benchmarks in this regard. Overall, we provide evidence that the QAOA, as applied to random Max-$k$XOR problems with odd and even $k$, might not show an advantage over classical algorithms.

\section*{Acknowledgments}
The authors acknowledge funding, support, and computational resources from the German Aerospace Center (DLR) and Forschungszentrum Jülich. Furthermore, we acknowledge funding from QSolid via the Federal Ministry of Education and Research (BMBF). We also acknowledge funding from AQUAS and QUASIM, both via the Federal Ministry of Economic Affairs and Climate Action (BMWK). Finally, we are grateful for useful conversations with Stuart Hadfield, Sevag Gharibian, Alessandro Ciani and Dmitry Bagrets.
   
\bibliographystyle{apsrev4-2}
\bibliography{ref} 
   	
\end{document}